\documentclass[floats,preprint,prb,aps,amssymb]{revtex4}
\usepackage{graphicx}
\DeclareGraphicsExtensions{.jpeg,.pdf,.eps,.gif,.tif,.sxd}
\DeclareGraphicsExtensions{.pdf}

\begin{document}
\title{Stochastic simulation of chemical exchange in two dimensional infrared spectroscopy}
\author{ Franti\v{s}ek \v{S}anda 
$^{* \dagger}$
 and \ Shaul \ Mukamel 
$^{*}$
}
\address{$^{*}$
Department of Chemistry, University of California, Irvine, CA 92697-2025\\
\\ $^{\dagger}$
Charles University, Faculty of Mathematics and Physics, Institute of Physics, Ke Karlovu 5, 
Prague, 121 16 The Czech Republic
}
\date{\today}
\begin{abstract}
\widetext
\bigskip
The stochastic Liouville equations are employed to investigate the combined  signatures of
chemical exchange (two-state-jump) and spectral diffusion  (coupling to an overdamped Brownian
oscillator) in  the coherent response of an anharmonic vibration to three femtosecond infrared pulses.
Simulations reproduce the main features recently observed in the OD stretch of phenol in benzene.

\vfill E-mail: smukamel@uci.edu
\end{abstract}
\maketitle
\section{Introduction}
Recent experiments had demonstrated that two-dimensional infrared (2D IR) lineshapes can probe
the picosecond  dynamics of chemical exchange by observing 
coherence transfer in molecular vibrations through time-dependent
spectral jumps \cite{fayer,fayer1,kim}. This
spectroscopic technique \cite{mukamel,jonas,multi} is an optical analogue of 2D NMR commonly
used to study slower (ms) chemical exchange processes  \cite{gamliel,kubo1,meier,ernst}.
In this  three-pulse experiment (Fig 1) the first pulse creates a vibrational
coherence, whose evolution (free induction 
 decay) during the first interval $t_1$ is related to the absorption lineshape by a Fourier transform.
During  the second interval $t_2$ the vibrational frequency changes by complexation with the solvent.
Finally vibrational  coherence is again created by the third pulse and detected during the third interval $t_3$. 
The correlations of the lineshapes in the first
and the third intervals provide information on chemical exchange taking place during the second interval.

One experiment \cite{fayer} looked at the OD stretching mode of phenol in benzene
whose frequency changes between
$ 2665 cm^{-1}$ (free)  and $2631 cm^{-1}$ (complexed). The 2D lineshapes  monitor the dynamics of
complex formation and dissociation \cite{fayer}. 
For short delays ($t_2 \sim 100 fs$) two diagonal peaks broadened by spectral diffusion were observed at the
linear-absorption frequencies.
 Off-diagonal cross-peaks
are induced by chemical exchange for longer times ($\sim 10 ps $).
Similar effects were seen in the overtone region. The rate constants for complexation 
were directly obtained from the peak kinetics.

 Similar hydrogen-bonding dynamics were also reported for the CN stretch 
of acetonitrile in methanol \cite{kim}.
The relative intensity of the two peaks attributed to free and hydrogen-bonded CN   
shows strong temperature dependence.
The reaction coordinate profile in  both ground and excited vibrational  states
was found to be very similar.
The weak variation of the chemical exchange rates with changes in the vibrational  state justifies
the application of stochastic chemical exchange models \cite{kampen} 
for vibration frequency fluctuations.   

In this paper we use the stochastic Liouville equations (SLE) to simulate 
the effects  of chemical exchange \cite{vinogradov,schuster,elsaesser} and spectral
diffusion in 2D lineshapes of an anharmonic vibration.
This formalism was recently applied to model the signatures of
conformational fluctuations in trialanine in water
\cite{jansen1} using
a Brownian oscillator coordinate model of fluctuations.
The SLE were also used to compare  the Brownian oscillator and the four-state-jump model of hydrogen bonding
fluctuation on the photon echo of the OH stretch of HOD in DOD \cite{jansen2}.

In section \ref{sec:model} we introduce the anharmonic vibrational
Hamiltionian subjected to fluctuations due to a two-state chemical-exchange
 and a Brownian oscillator coordinate.
In the  two state jump (TSJ) model of Kubo and Anderson \cite{kubo,kubo1,anderson,zhao,sanda}
bath-assisted (incoherent) chemical exchange is described by kinetic equations.
The model shows several distinct dynamical regimes depending 
on the motional narrowing parameter
(ratio of the frequency shift and relaxation rate). 
Spectral diffusion \cite{fayer} is incorporated 
by coupling to an overdamped Brownian oscillator
coordinate \cite{legget}. 
The 2D IR signals are defined in section \ref{sec:response}.
The stochastic Liouville equation for the
TSJ model is introduced in section \ref{sec:SLE} and the 2D IR signals 
are calculated for the limiting cases of  slow and fast fluctuations
and for short and long-time delays between pulses. The combined effect of
spectral diffusion and TSJ is presented and analyzed  in section
\ref{sec:spectral_diffusion}.
All lineshape regimes observed in \cite{fayer} are qualitatively reproduced.

\section{the Model}
\label{sec:model}
We consider a single anharmonic vibrational mode described by the Hamiltonian:
\begin{equation}
\label{hamiltonian}
H=\hbar\Omega B^{\dagger}B+\hbar\frac{\Delta}{2}  B^{\dagger} B^{\dagger}B B
\end{equation}
where $B^{\dagger}$ ($B$) are boson creation (annihilation) operators ($\left[B,B^{\dagger}\right]=1$).
Both frequency ($\Omega$) and anharmonicity ($\Delta$) are subjected to fluctuations described by
\[
\Omega=\Omega_0+\Omega_1 \sigma_z +\Omega_2 Q+ \Omega_3 \sigma_z Q,
\]
\begin{equation}
\Delta=\Delta_0+\Delta_1 \sigma_z +\Delta_2 Q +\Delta_3 \sigma_z Q.
\label{fluctuations}
\end{equation}
 $\Omega_0$ and $\Delta_0$ are the average values. $\Omega_1$  and $\Delta_1$ describe stochastic frequency modulation by
chemical exchange represented by the TSJ model  \cite{kubo,anderson}
(state up $u$ ($\sigma_z=1$) and down $d$ ($\sigma_z=-1$)). $\sigma_z$ is the Pauli spin matrix .
\[
\sigma_z=\left(\begin{array}{cc}1&0 \\
0&-1 \end{array} \right)
\]
 $Q$ is an
overdamped Brownian oscillator (BO) coordinate causing
spectral diffusion and coupled to the vibration through the parameters $\Omega_2$, and $\Delta_2$.
$\Omega_3$, and $\Delta_3$  allow the spectral diffusion 
to have a different magnitude in the two spin states.
We shall divide  hamiltonian (Eq. (\ref{hamiltonian})) into three parts  $H=H_{SV}+H_{QV}+H_{QSV}$
where $H_{SV}$ includes the first two (spin-vibration) terms in Eq.(\ref{fluctuations}) which do not depends on $Q$,
and $H_{QV}$,$H_{QSV}$ include the third and fourth terms in Eq.(\ref{fluctuations}) respectively.
The dipole interaction  with the electric field $E(t)$ is \cite{cohen}
\begin{equation}
H_{int}=-\mu E(t)\left( B^{\dagger}+B \right) 
\label{interaction}
\end{equation}
Three vibrational  levels (Fig 1) are accessible in a three-pulse experiment: 
the ground state  $|g\rangle$, the first excited $|e\rangle\equiv B^{\dagger} |g\rangle$ and
 the doubly excited $|f\rangle\equiv(1/\sqrt{2})B^{\dagger} B^{\dagger} |g\rangle$ state,
with energies 0, $\Omega$ and $\Omega+\Delta$ respectively.
The corresponding 9-component Liouville space basis set for the density matrix is denoted 
$|\nu\nu'\rangle \rangle \equiv |\nu \rangle \langle \nu'|; \quad \nu,\nu'=e,f,g$.
The dipole moment matrix elements are $\mu_{eg} =\mu$;  $\mu_{ef}=\sqrt{2}\mu$.

\section{The Third order response and  2D signals }
\label{sec:response}

We consider an impulsive four wave mixing process whereby the vibrational mode (Eq.(\ref{hamiltonian}))
interacts with three impulsive optical pulses with intervals $t_1,t_2$, and
calculate the response at $t_3$ (see Fig 1). Each interaction
with the field can create or annihilate one vibrational quantum at a time
(Eq(\ref{interaction})). 
The equilibrium distribution is
\begin{equation}
|\rho(0) \rangle \rangle_{\mathcal H}=|\rho(0)\rangle_S  |0\rangle_Q |gg\rangle \rangle
\label{equilibrium2}
\end{equation}
where $|\rho(0)\rangle_S $ is equilibrium spin state  and $|0\rangle_Q$ equilibrium Brownian coordinate density 
given by Eq. (\ref{herm}).
The space  ${\mathcal H}$ is
a direct product of the vibrational, spin and BO space.

The polarization generated in this experiment is described by the third order
response function \cite{mukamel}
\begin{equation}
S^{(3)}(t_1,t_2,t_3)=\left(\frac{i}{\hbar}\right)^3\theta (t_1) \theta (t_2) \theta (t_3)
 \langle \langle I| \mu^{(-)}
{\mathcal G}(t_3)  \mu^{(-)} {\mathcal G}(t_2)  \mu^{(-)} {\mathcal G}(t_1)  \mu^{(-)} |\rho(0)\rangle \rangle_{\mathcal H}
\end{equation}
where $\mu^{(-)}\xi \equiv \mu [B+B^{\dagger},\xi ]$.
Summing over final states is represented by $\langle\langle  I|_{\mathcal H}\equiv  \langle I|_S \langle 0|_Q Tr$ where
$\langle  I|_S\equiv (1,1)$, $\langle 0|_Q\equiv \int dQ$  and tracing is over the vibrational degrees of freedom 
$Tr=\langle \langle gg|+ \langle \langle ee| +\langle \langle ff|$.
The time evolution between pulses is described by the Green's function ${\mathcal G}(t)$ of the SLE
defined  in the space  ${\mathcal H}$.

The Liouville space state cannot change between pulses, so that 
Green functions are block-diagonal.
The response functions can be separated
into six contributions from different Liouville space pathways. 
\begin{eqnarray}
R_i(t_3,t_2,t_1)&\equiv& \langle I|
{\mathcal G}_{eg,eg}(t_3)  {\mathcal G}_{ee,ee}(t_2) {\mathcal G}_{eg,eg}(t_1)  |\rho(0)\rangle_{QS}
\nonumber \\
R_{ii}(t_3,t_2,t_1)&\equiv&\langle  I|
{\mathcal G}_{eg,eg}(t_3)  {\mathcal G}_{ee,ee}(t_2) {\mathcal G}_{ge,ge}(t_1)  |\rho(0)\rangle_{QS}
\nonumber \\
R_{iii}(t_3,t_2,t_1)&\equiv& \langle I|
{\mathcal G}_{eg,eg}(t_3)  {\mathcal G}_{gg,gg}(t_2) {\mathcal G}_{eg,eg}(t_1)  |\rho(0)\rangle_{QS}
\nonumber \\
R_{iv}(t_3,t_2,t_1)&\equiv&\langle  I|
{\mathcal G}_{eg,eg}(t_3)  {\mathcal G}_{gg,gg}(t_2) {\mathcal G}_{ge,ge}(t_1)  |\rho(0)\rangle_{QS}
\nonumber \\
R_{v}(t_3,t_2,t_1)&\equiv&-\langle I|
{\mathcal G}_{fe,fe}(t_3)  {\mathcal G}_{ee,ee}(t_2) {\mathcal G}_{eg,eg}(t_1)  |\rho(0)\rangle_{QS}
\nonumber \\
R_{vi}(t_3,t_2,t_1)&\equiv& - \langle I|
{\mathcal G}_{fe,fe}(t_3)  {\mathcal G}_{ee,ee}(t_2) {\mathcal G}_{ge,ge}(t_1)  |\rho(0)\rangle_{QS}
\label{resdef}
\end{eqnarray}
 These 
may be graphically represented by the Feynman diagrams \cite{principles} given in Fig 1.
For the stochastic models of frequency fluctuations  considered here
we have $R_{i}=R_{iii}$ and $R_{ii}=R_{iv}$, so that
we only need to consider four independent pathways \cite{principles}.
The Green's function matrix elements ${\mathcal G}_{fe,fe}(t_3)$ etc. are now matrices
 in the joint spin and BO space.

We shall display the signal using the  mixed time-frequency representation 
\begin{equation}
R_{\alpha}(\omega_3,t_2,\omega_1)\equiv\int_0^{\infty}
\int_0^{\infty} \exp{\left(i\omega_1t_1+i\omega_3t_3\right)}
R_{\alpha}(t_3,t_2,t_1) d t_3 d t_1
\label{mixdomain}
\end{equation}
Eq. (\ref{resdef}) then reads
\[
R_i(\omega_3,t_2,\omega_1)=\langle I| {\mathcal
G}_{eg,eg}(\omega_3)  {\mathcal G}_{ee,ee}(t_2) {\mathcal
G}_{eg,eg}(\omega_1)  |\rho(0) \rangle_{QS} ;
\]
and similarly for other pathways , where 
\[
{\mathcal G}(\omega)\equiv  \int_0^{\infty} {\mathcal G}(t) e^{i\omega t} d t
\]

We consider the  coherent nonlinear signals
generated in the ${\bf k_I=-k_1+k_2+k_3}$ and ${\bf k_{II}=k_1-k_2+k_3}$ phase-matching  directions.
Invoking the rotating wave approximation the ${\bf k_I}$ (photon echo) signal  is \cite{principles}
\begin{equation}
S_I(\omega_3,t_2,\omega_1)= 
 \left(\frac{i}{\hbar}\right)^3
\left\{\mu_{eg}^4 \left[ R_{ii}(\omega_3,t_2,\omega_1)
+R_{iv}(\omega_3,t_2,\omega_1)\right] +\mu_{eg}^2\mu_{ef}^2
R_{vi}(\omega_3,t_2,\omega_1) \right\} 
\label{k_1}
\end{equation}
and for the ${\bf k_{II}}$ signal we have
\begin{equation}
S_{II}(\omega_3,t_2,\omega_1)= 
 \left(\frac{i}{\hbar}\right)^3
\left\{\mu_{eg}^4 \left[ R_i (\omega_3,t_2,\omega_1)
+R_{iii}(\omega_3,t_2,\omega_1)\right] +\mu_{eg}^2\mu_{ef}^2
R_{v}(\omega_3,t_2,\omega_1) \right\}
 \label{k_2}
\end{equation}
We shall also display the following combination which shows  absorptive peaks
\cite{tokmakoff,scheurer}.
\begin{equation}
S_A(\omega_3,t_2,\omega_1) \equiv -Im \left[S_I(\omega_3,t_2,-\omega_1) +S_{II}(\omega_3,t_2,\omega_1) \right]
\label{2DIRdef}
\end{equation}

When during the interval $t_3$ the system has no memory about its state in $t_1$ 
the third order response functions are factorized  into products of the linear response functions $K(t)$
\[
R_{i}(t_3,t_2,t_1)= K(t_3) K(t_1)
\]
\begin{equation}
R_{ii}(t_3,t_2,t_1)=K(t_3) K^*(t_1); \quad \quad R_{ii}(\omega_3,t_2,\omega_1)=K(\omega_3) K^*(-\omega_1);
\label{prefactorization}
\end{equation}
Neglecting the overtone contributions $R_{v}$ and $R_{vi}$  it follows
that the $S_A$ signal is given by  the product of the linear lineshapes $W_A$ 
\begin{equation}
\hbar S_A(\omega_3,\omega_1)= W_A(\omega_1) W_A(\omega_3).
\label{factorization}
\end{equation}

\section{Stochastic Liouville equations for the two-state-jump model}
\label{sec:SLE}

We first consider the TSJ model \cite{kubo,kubo1,anderson} 
by neglecting the spectral diffusion (setting $\Delta_{2}, \Delta_3,\Omega_{2}, \Omega_3=0$).
The total density matrix $\rho$ has 18 components  $|\nu \nu' s\rangle\rangle$ given by the 
direct product of 9
Liouville space  states $|\nu \nu' \rangle\rangle$ and the two spin states, $\quad s=u,d$.
The SLE \cite{kubo3,kampen} then reads
\begin{equation}
\frac{d\rho}{dt}=\hat{\mathcal L}\rho(t).
\label{SLE}
\end{equation}
The Liouville operator $\hat{\mathcal L}$ is diagonal in the vibrational Liouville space variables,
and is thus given by nine $2\times 2$ diagonal blocks in spin space,
\begin{equation}
\left[\hat{\mathcal L}\right]_{\nu\nu's,\nu_1\nu_1's'}=
\delta_{\nu \nu_1} \delta_{\nu' \nu_1'} \left[\hat{\mathcal L}_S\right]_{s,s' } +
\delta_{\nu \nu_1} \delta_{\nu' \nu_1'}\delta_{ss'} \left[\hat{\mathcal L}_{SV}\right]_{\nu \nu's,\nu \nu' s }
\label{TSJ_SLE}
\end{equation}
where $\hat{\mathcal L}_S$ describes the two state jump kinetics and
 $\hat{\mathcal L}_{SV}$ represents the coherent vibrational evolution that depends parametricaly on the spin state.

Chemical exchange is described by the rate matrix
$\hat{\mathcal L}_S$ for the spin:
\begin{equation}
\left[\hat{\mathcal L}_S\right]=
\left(\begin{array}{cc}-k_d &k_u \\
k_d&-k_u \end{array} \right)
\label{spin}
\end{equation}
The up (down) jump rates $k_u$, ($k_d$) 
are connected by the detailed-balance relation $ k_u
/ k_d =\exp{\beta (\epsilon_d-\epsilon_u)}$, where
$\epsilon_d-\epsilon_u$ is the energy difference between the $u$
and $d$ states.

The equilibrium density matrix is
\begin{equation}
 |\rho(0)\rangle\rangle=|gg\rangle\rangle |\rho(0)\rangle_S;
\quad |\rho(0)\rangle_S =\frac{1}{k_u+k_d}\left(\begin{array}{cc}k_u\\
k_d\end{array} \right).
\label{equilibrium1}
\end{equation}

We next turn to the vibrational part $\hat{\mathcal L}_{SV}\equiv -(i/\hbar)[H_{SV},\ldots]$.
For the  $|gg \rangle\rangle$,  $|ee \rangle\rangle$,  $|ff \rangle\rangle$ blocks
$\left[\hat{\mathcal L}_{SV}\right]_{gg,gg}=
\left[\hat{\mathcal L}_{SV}\right]_{ee,ee}=
\left[\hat{\mathcal L}_{SV}\right]_{ff,ff}=0$.
The other blocks of ${\mathcal L}_{SV}$ are:
\[
\left[\hat{\mathcal L}_{SV}\right]_{eg,eg}=\left(\begin{array}{cc} -i(\Omega_0 +\Omega_1) &0 \\
0&-i (\Omega_0 -\Omega_1)\end{array} \right)
\]
\[
\left[\hat{\mathcal L}_{SV}\right]_{fe,fe}=\left(\begin{array}{cc}-i(\Omega_0+\Delta_0)-i(\Omega_1+\Delta_1) &0 \\
0&-i (\Omega_0+\Delta_0) +i(\Omega_1+\Delta_1)\end{array} \right)
\]
\[
\left[\hat{\mathcal L}_{V}\right]_{fg,fg}=\left(\begin{array}{cc}-i(2\Omega_0+\Delta_0)-i(2\Omega_1+\Delta_1) &0 \\
0&-i (2\Omega_0+\Delta_0) +i(2\Omega_1+\Delta_1)\end{array} \right)
\]
The remaining blocks are obtained by taking complex conjugates
$\left[\hat{\mathcal L}_{SV}\right]_{\nu \nu',\nu \nu' }=\left[\hat{\mathcal L}_{SV}\right]^*_{\nu' \nu ,\nu'\nu }$.

These matrices may be represented in a compact form by defining the energy
$\hbar \epsilon_0$ and splitting $\hbar \epsilon_1$ parameters
$\epsilon_0^{(g)}=0$, $\epsilon_0^{(e)}=\Omega_0$, $\epsilon_0^{(f)}=2\Omega_0+\Delta_0$,
$\epsilon_1^{(g)}=0$, $\epsilon_1^{(e)}=\Omega_1$, $\epsilon_1^{(f)}=2\Omega_1+\Delta_1$.
We then have
\begin{equation}
\left[\hat{\mathcal L}_{SV}\right]_{\nu \nu',\nu_1\nu_1' }=\delta_{\nu \nu_1} \delta_{\nu' \nu_1'}
\left(\begin{array}{cc}-i(\epsilon_0^{(\nu )}-\epsilon_0^{(\nu')}) -i(\epsilon_1^{(\nu )}-\epsilon_1^{(\nu')})
&0 \\
0& -i(\epsilon_0^{(\nu )}-\epsilon_0^{(\nu')}) +i(\epsilon_1^{(\nu )}-\epsilon_1^{(\nu')})
\end{array} \right)
\label{spindephasing1}
\end{equation}

The Green's function solution of Eq. (\ref{SLE}) is given by $2\times 2$ blocks 
for each vibrational state $|\nu \nu'\rangle \rangle$ of the density matrix 
\[
\left[{\mathcal G}\right]_{\nu \nu',\nu_1\nu_1' }
 (t)\equiv\left[\exp{\left(\hat{\mathcal L}t\right)}\right]_{\nu \nu',\nu_1\nu_1' }=\delta_{\nu \nu_1} \delta_{\nu' \nu_1'}
\]
\begin{equation}
\times \left\{\left(\frac{\eta_2}{\eta_2-\eta_1}\hat{1}
-\frac{1}{\eta_2-\eta_1}\hat{\mathcal L}_{\nu \nu',\nu \nu'} \right) \exp{(\eta_1 t)}
+\left(\frac{\eta_1}{\eta_1-\eta_2}\hat{1}
-\frac{1}{\eta_1-\eta_2}\hat{\mathcal L}_{\nu \nu',\nu \nu'} \right) \exp{(\eta_2 t)}
\right\}
\label{green_function}
\end{equation}
where $\eta_{j}$ are the eigenvalues of each block of $\hat{\mathcal L}$
\[
\eta_{1}=-\frac{k_d+k_u}{2}- i(\epsilon_{0}^{(\nu)}-\epsilon_{0}^{(\nu')})
+\sqrt{
\frac{(k_d+k_u)^2}{4} -
(\epsilon_{1}^{(\nu)}-\epsilon_{1}^{(\nu')})^2
+ i(\epsilon_{1}^{(\nu)}-\epsilon_{1}^{(\nu')})(k_d-k_u)
}
\]
\[
\eta_{2}=-\frac{k_d+k_u}{2}- i(\epsilon_{0}^{(\nu)}-\epsilon_{0}^{(\nu')})
-\sqrt{
\frac{(k_d+k_u)^2}{4} -
(\epsilon_{1}^{(\nu)}-\epsilon_{1}^{(\nu')})^2
+ i(\epsilon_{1}^{(\nu)}-\epsilon_{1}^{(\nu')})(k_d-k_u)
}
\]

For the  $gg$,$ee$ and $ff$ space $\hat{\mathcal L}_{SV}=0$,
 and Eq.(\ref{green_function}) reads
\begin{equation}
{\mathcal G}_{ee,ee}(t)={\mathcal G}_{gg,gg}(t)={\mathcal G}_{ff,ff}(t)\equiv
\exp{\left(\hat{\mathcal L}_St\right)}=
\hat{1}+\frac{1-\exp{[-(k_d+k_u)t]}}{k_d+k_u}
\left(\begin{array}{cc}-k_d  &k_u \\
k_d&-k_u \end{array} \right)
\label{decoupled}
\end{equation}
The linear
response is given in Appendix \ref{sec:Linear_response}.
Closed form expressions for the various pathways are given
in the Appendix \ref{sec:kubo_anderson_response}.
Below we discuss the 2D signals for limiting cases.

In the slow bath limit $\Omega_1>>k$  (In Figs 2-6 $k\equiv k_d=k_u$) 
no spin jumps occur during the  $t_1$,and $t_3$ intervals.
In Fig 2a we display the 2D lineshapes $S_A$ (Eq.(\ref{2DIRdef})). 
In all figures we give frequencies with respect to $\Omega_0$
(i.e. we set  $\Omega_0=0$) and normalize the signal
to have maximum absolute value of 1. 
For $ kt <<1$ no spin jumps occur during the  $t_2$ interval and
 we only see
4 peaks (Fig 2a) two diagonal  at $\omega_{1}=\omega_{3}=\Omega_0\pm \Omega_1$
and and two overtone at $(\omega_1,\omega_3)=(\Omega_0+\Omega_1,\Omega_0+\Delta_0 +\Omega_1+\Delta_1)$
and $ (\Omega_0-\Omega_1,\Omega_0+\Delta_0 -\Omega_1-\Delta_1)$
In the opposite $k t_2 >>1$ limit we
see  8 peaks (Fig 2b). In addition to the previous four peaks cross peaks appears at 
$(\omega_1,\omega_3)=(\Omega_0+\Delta_0,\Omega_0-\Delta_0), (\Omega_0-\Delta_0, \Omega_0+\Delta_0),
(\Omega_0-\Omega_1,\Omega_0+\Delta_0+\Omega_1+\Delta_0) , (\Omega_0+\Omega_1,\Omega_0-\Omega_1+\Delta_0-\Delta_1)$
 due to the spin jumps. Diagrams $i$, $ii$, $iii$, and $iv$ describe the $g$ to $e$ peaks
at $\omega_1, \omega_3\sim \Omega$. Diagrams $v$ and $vi$ show the
$e$ to $f$ peaks at $\omega_1 \sim \Omega+\Delta$.

Note that the relation (Eq.(\ref{slow_peak_shapes})), similar to  Eq.(\ref{factorization}) is valid
for {\it individual} peaks in the short time regime since peak shapes are not connected with 
any inhomogenity which carries memory, 
however  Eq.(\ref{factorization}) does not apply for the whole 2D shape, since the memory of spin state erases the cross peaks.
In contrast, for long $t_2$  (Fig 2b) this memory is lost
and Eq. (\ref{factorization}) applies (excluding the overtone contributions).

In Fig 2c we plot $S_A$ (Eq.\ref{2DIRdef}) in the fast $k_u,k_d>>\Omega_1$ limit.
We see a fundamental peak at $\omega_{1}=\omega_{3}=\Omega_0$ and
an overtone peak at $\omega_1=\Omega_1$, $\omega_3=\Omega_0+\Delta_0$.
Linear response in fast regime shows one peak (fluctuational narrowing \cite{kubo}). 
Any memory of $t_1$ interval is 
lost in the $t_3$ interval and the factorization  (Eq.\ref{prefactorization}) is valid
resulting in Eq. \ref{factorization} which is observed in Fig 2c.

\section{Two-state-jump with spectral diffusion}
\label{sec:spectral_diffusion}
Spectral diffusion is described by
a dimensionless  overdamped Brownian oscillator coordinate $Q$ (Eq.(\ref{hamiltonian}))
whose variance $1$ and its dynamics is given by the Fokker-Planck operator \cite{risken}
\begin{equation}
\label{fp}
{\mathcal L}_Q = \Lambda \frac{\partial}{\partial Q}\left(Q+\frac{\partial}{\partial Q}\right)
\end{equation}
where $\Lambda$ is the relaxation rate.
For fast fluctuations $\Omega_2/\Lambda<<1$, 
spectral diffusion may be accounted for by simply adding dephasing
rates $\Omega_2^2/\Lambda $ to the lineshapes.
The complete solution of the Fokker-Planck equation is required for arbitrary fluctuation timescales.

We define $\epsilon_2^{(g)}=0$, $\epsilon_2^{(e)}=\Omega_2$, $\epsilon_2^{(f)}=2\Omega_2+\Delta_2$, 
and $\epsilon_3^{(g)}=0$, $\epsilon_3^{(e)}=\Omega_3$, $\epsilon_3^{(f)}=2\Omega_3+\Delta_3$
The coupling of $Q$ to the vibration is given by the Liouville operator $\hat{\mathcal L}_{QV}$
\begin{eqnarray}
\left[\hat{\mathcal L}_{QV} \right]_{\nu \nu' s, \nu_1\nu_1's_1}=
&-&i\left(\epsilon_2^{(\nu)}-\epsilon_2^{(\nu')}\right)Q\delta_{\nu\nu_1}\delta_{\nu'\nu_1'}\delta_{ss_1}
\nonumber \\
\left[\hat{\mathcal L}_{SQV} \right]_{\nu \nu' s, \nu_1\nu_1's_1}
&-&i\left(\epsilon_3^{(\nu)}-\epsilon_3^{(\nu')}\right)Q\delta_{\nu\nu_1}\delta_{\nu'\nu_1'}\left[\sigma_z\right]_{ss_1}
\label{Qdephasing1}
\end{eqnarray}

The complete Liouville superoperator describing both TSJ and BO fluctuations is finally  given by:
\begin{equation}
\hat{\mathcal L}_{tot}=\hat{\mathcal L}_S+\hat{\mathcal L}_{SV}+\hat{\mathcal L}_{QV}+ \hat{\mathcal L}_{SQV}+{\mathcal L}_Q
\label{totalSLE}
\end{equation}
where $\hat{\mathcal L}_S$ is given by Eq. (\ref{spin}), $\hat{\mathcal L}_{SV}$ by Eqs.(\ref{spindephasing1}),
$\hat{\mathcal L}_{QV}$,$\hat{\mathcal L}_{QSV}$ by Eqs.(\ref{Qdephasing1}), and ${\mathcal L}_Q$ by Eq.(\ref{fp}).

The response function for the spectral diffusion model alone \cite{tanimura2}
(neglecting the spin, setting $\Omega_1=0$,$\Delta_1=0$) can be
calculated using the second order cumulant expansion
\cite{principles} and is given in Appendix \ref{sec:Cumulant}.

We next turn to  
the third order response (Eq.(\ref{resdef})). The
equilibrium distribution is
\begin{equation}
|\rho(0) \rangle_{\mathcal QS}=|\rho(0)\rangle_S  |0\rangle_Q
\label{equilibrium3}
\end{equation}
where $|\rho(0)\rangle_S $ is given by Eq.(\ref{equilibrium1}) and $|0\rangle_Q$ by Eq. (\ref{herm})
and the final averaging should run over all degrees of freedom.
\[
\langle I|_{\mathcal QS}=\langle I|_S\langle 0|_Q
\]
The Green's function matrix elements ${\mathcal G}_{fe,fe}(t_3)$ are now matrices
 in the joint spin and BO space
\begin{equation}
{\mathcal G}_{\nu \nu',\nu_1 \nu_1'}(\omega)
=-\left( \left[i\omega+ \hat{\mathcal L}_{tot}\right]_{\nu \nu',\nu_1 \nu_1'}\right)^{-1}.
\label{Green_function_frequency}
\end{equation}
${\mathcal G}(\omega_1)$, ${\mathcal G}(\omega_3)$ are calculated in  Appendix \ref{sec:Continued_fraction}
  by expanding them  in the Fokker-Planck eigenmodes \cite{jansen1,jansen2,risken,tanimura1}.

During  $t_2$ the evolution is in the  $ee,gg,ff$ space where $\hat{\mathcal L}_{QS}=0$
and the Green's function may be  factorized as
\begin{equation}
\label{separation}
{\mathcal G}_{ee,ee}(t_2)={\mathcal G}_{gg,gg}(t_2)=
\exp{\left(\hat{\mathcal L}_St_2\right)} \exp{\left({\mathcal L}_Qt_2\right)}
\end{equation}
where the first factor is given by Eq.(\ref{decoupled}) and the
second term is the propagator of Fokker-Planck equation
\cite{risken} expanded in its eigenmodes (Eq.(\ref{herm}))
\begin{equation}
\left[\exp{\left({\mathcal L}_Q t \right)}\right]_{\alpha \beta}= 
\delta_{\alpha \beta} \exp{\left(-\alpha \Lambda t\right)}
\label{fpev}
\end{equation}
For slow chemical exchange compared to the spectral diffusion ($k_u, k_d<<\sigma,\Lambda$) 
the TSJ peaks are well resolved and their lineshapes are determined by the spectral diffusion.
Chemical exchange occurs on a much longer timescale  and affects the cross peaks.  

We start by setting  $\Omega_3, \Delta_3=0$ and
examine  fast and the slow spectral diffusion.   
The fast (motional narrowing) limit ($\Lambda>>\Omega_2$) gives a
Lorentzian absorption lineshape.  Since the bath has no memory, peak shapes are less sensitive to $t_2$ 
and cross peaks due to  chemical exchange only appear as $t_2$ is increased.
Effects of fast fluctuations may be accounted for  by adding dephasing rates $\Omega_2^2/\Lambda$.
2D-lineshapes $S_A$ (Eq. (\ref{2DIRdef}) shown in Fig 3 are similar to  the TSJ model (Fig 2),
both are given by  a product of  Lorentzian lineshapes (in $\omega_1,\omega_3$).
The important difference is that when accounting  for fast spectral diffusion,
the linewidth is determined by both exchange
and dephasing rates $(k_d+\Omega_2^2/\Lambda)^{-1}$, i.e. it is different from the cross-peak timescale. 
In contrast the linewidth of TSJ model  must be equal to exchange rate $k$ obtained by analysis
of the cross-peaks growth.

$S_A$ in the static $\Lambda<<\Omega_2$ limit (Fig 4) shows Gaussian linear lineshapes. Static disorder is eliminated 
in the photon echo ${\bf k_I}$ experiment. 2D peaks are elliptic with different 
"diagonal" and "off-diagonal" linewidths (Fig 4a) \cite{okomura}.
The intermediate regime may be observed at $\Lambda ^{-1} <t_2< k^{-1}$ when the bath loses memory  (Fig 4b)
and the 2D peaks become symmetric. For long times $k_d t_2 >>1$ we see cross peaks induced by chemical exchange.  
Note the circular contours of Fig 4c, compared to the star-like contours for fast fluctuations (Fig 3).
These may be understood by noting  that the product of two Gaussians (which represent the linear lineshape for the slow case) 
is rotationally invariant, unlike the Lorentzian lineshapes in the fast limit.  
 
Fig 5 shows the ${\bf k_I}$ and ${\bf k_{II}}$ signals  (Eqs. (\ref{k_1}), and (\ref{k_2})) .
The ${\bf k_I}$ contribution dominates at short times, and shows  effective  rephasing (photon echo) and
the final 2DIR is similar to this dominating ${\bf k_I}$ contribution (See Fig 4a,).  
The bath loses memory of its initial state in the third interval. For longer $t_2>>\Lambda^{-1}$
(right panels), the ${\bf k_I}$ and ${\bf k_{II}}$ contributions become comparable and $S_A$
 becomes symmetric (see Fig 4). 
 
We next explore some more general cases. First we take $\Delta_2 \neq 0$.
The overtone peaks may then show different widths along $\omega_3$ and $\omega_1$. 
(overtone photon echoes are expected at $\Omega_2 t_1=(\Omega_2+\Delta_2)t_3$ )
Note that we have a
slow bath ($\Lambda<\Omega_2$) for the $eg$ and a fast bath for the $ef$ transition ($\Lambda > \Omega_2+\Delta_2$).
These lineshapes are shown in Fig. 6. 

The inclusion of $\Omega_3$ does not require additional numerical effort.
This parameter allows the $u$ and $d$ peaks have a different width due 
the spectral diffusion.

To demonstrate these effects and relate them to  recent experiments \cite{fayer} 
we have simulated the spectral diffusion using a single overdamped Brownian coordinate
allowing a different width for $u$ and $d$ peaks.
We used the splitting  $2\Delta_0=34cm^{-1}$
 (i.e. $\sim 1.01 ps^{-1}$) and the exchange rates $k_u=0.1 ps^{-1}$ $k_d=0.125 ps^{-1}$  which were
 determined in \cite{fayer} from the cross peaks growth.
 Three regimes similar to those of Fig 4 were observed \cite{fayer} suggesting that the bath is not fast.
 The intermediate regime, where the memory of the BO coordinate is
 lost (circular lineshape) but the cross peaks are still weak, 
  was found on the  2ps timescale. We thus assumed for the relaxation rate  $\Lambda \sim 0.4 ps^{-1}$.
 Knowing $\Lambda$, $\Omega_1$ and $\Omega_3$ are connected to the peak linewidth in linear spectra and
can be estimated using the Pade approximant of a 2-level system \cite{yan,principles}. 
We have simulated the absorption spectra in Fig 7 and found  lineshapes similar to experiment
for $\Omega_2=0.33ps^{-1}$, $\Omega_3=-0.07ps^{-1}$. 
This completely determines the model for 2D IR spectroscopy (neglecting dynamics in the overtone),
we have no additional free parameter.
In Fig 18 we show that the predicted  2DIR lineshapes of the same model
 are in qualitative agreement  with  experiment.
We see all  three regimes, rephasing eliptic shapes , the relaxed Brownian oscillator with circular
shape  and chemical exchange cross-peaks  at the proper timescales. 
  The peaks have also the correct relative intensity the lower frequency peak is weaker but broader. 
 
In conclusion, we have demonstrated that the SLE may be used
to model chemical exchange in 2DIR spectroscopy.
Spectral diffusion can no longer be accounted for in terms of 
dephasing rates, when its timescale may be observed in experiment and
its proper description must be combined 
with the two state jump which describe chemical exchange.
The high temperature  overdamped Brownian oscillator  model for  
spectral diffusion with arbitrary fluctuation timescale 
reproduces the most significant features of recent experiments \cite{fayer}.

In contrast  with  calculations based on cumulant expansions
at finite temperatures \cite{jansen2,principles,markham}
our high temperature bath does not respond to the state of system, 
and its evolution in the ground and excited state is same.
 This may limit the applicability of SLE model when signatures of the vibrational Stokes shift are observed in 2DIR signals, 
however it simplifies the calculations.
The lack of phase factors during the $t_2$ interval reduces the dynamics to classical level 
and allows large scaleMD simulations of
environmental dynamics.

\acknowledgements The support of NSF (Grant No CHE-0446555) and
NIH (2RO1GM59230-05) is gratefully acknowledged. F. \v{S}. is also supported by the Ministry of Education,
Youth and Sports of the Czech Republic (project MSM 0021620835).
\appendix

\section{the Absorption lineshape}

\label{sec:Linear_response}
The linear response function is given by \cite{principles}
\[
 S^{(1)}(t_1)=\theta(t_1)\frac{i}{\hbar}\langle\langle I|\mu^{(-)}{\mathcal G}(t_1) \mu^{(-)}|\rho(0)\rangle\rangle_{\mathcal H}
\]

We define the contribution
\[
 K(t_1)=\langle I|{\mathcal G}_{eg,eg}(t_1) |\rho(0)\rangle_{QS}.
\]

This response function is connected to absorption lineshape
\begin{equation}
W(\omega_1)= \mu_{eg}^2 \hbar^{-1} Re
\int_0^{\infty}\exp{\left(i\omega_1 t_1\right)} K(t_1) dt_1
\label{linear_lineshape}
\end{equation}

For the TSJ model ($\Omega_{2,3}=0$) the
result of Kubo \cite{kubo} is recovered:
\[
W(\omega) =\frac{1}{(k_d+ k_u)}\times\frac{4 k_d k_u\Omega_1^2}{(\omega-\Omega_0-\Omega_1)^2
+\left[ (\omega -\Omega_0)(k_d+k_u)+\Omega_1(k_d-k_u)\right]^2
}
\]

For the spectral diffusion model we set $\Omega_1=0$ and  the linear response
function is given by the second order cumulant expression (see Eq. (\ref{linebroad})).
\[
K(t_1)= \exp{\left(-g_{ee}(t_1)\right)}
\]

\section{Response functions for  two-state-jump spin dynamics}
\label{sec:kubo_anderson_response}
In the frequency domain the Green's function (Eq. \ref{green_function}) matrix elements
are:
\[
{\mathcal G}_{\nu \nu',\nu_1 \nu_1'}(\omega)= -\left[\left(i\omega+ \hat{\mathcal L}\right)^{-1}\right]_{\nu \nu',\nu_1 \nu_1'}
=
\]
\[
=
\frac{-\delta_{\nu \nu_1}\delta_{\nu' \nu_1'}}{(\omega-\epsilon_{0}^{(\nu)}+\epsilon_{0}^{(\nu')})^2-
(\epsilon_{1}^{(\nu)}-\epsilon_{1}^{(\nu')})^2
+i(\omega-\epsilon_{0}^{(\nu)}+\epsilon_{0}^{(\nu')})(k_d+k_u)
+i(\epsilon_{1}^{(\nu)}-\epsilon_{1}^{(\nu')})(k_d-k_u)}
\]
\begin{equation}
\times\left(\begin{array}{cc}k_u -i(\omega-\epsilon_{0}^{(\nu)}+\epsilon_{0}^{(\nu')}
+\epsilon_{1}^{(\nu)}-\epsilon_{1}^{(\nu')})&k_u \\
k_d&k_d -i(\omega-\epsilon_{0}^{(\nu)}+\epsilon_{0}^{(\nu')}
-\epsilon_{1}^{(\nu)}+\epsilon_{1}^{(\nu')})
\end{array} \right)
\end{equation}

$R_{i}$ and $R_{ii}$ (Eq.\ref{resdef}) are given by:
\[
R_{\alpha}(\omega_3,t_2,\omega_1)=
\frac{1}{(\omega_1\mp\Omega_0)^2-\Omega_{1}^2+i(\omega_1\mp\Omega_{0})(k_d+k_u)\pm i\Omega_{1}(k_d-k_u)}
\]
\[
\times \frac{1}{(\omega_3-\Omega_{0})^2-\Omega_{1}^2+i(\omega_3-\Omega_{0})(k_d+k_u)+ i\Omega_{1}(k_d-k_u)}
\frac{1}{k_u+k_d}
\]

\[
\times \bigg\{(k_u+k_d)^3-\left[(\omega_3-\Omega_{0})(\omega_1\mp\Omega_{0})\pm\Omega_{1}^2\right](k_u+k_d)
+\left[(\omega_1\mp\Omega_{0})\Omega_{1}\pm(\omega_3-\Omega_{0})\Omega_{1}\right](k_d-k_u)
\]
\[
-i\left[(\omega_3-\Omega_{0}+\omega_1\mp\Omega_{0})(k_u+k_d)^2
-(\Omega_{1}\pm \Omega_{1})(k_d^2-k_u^2)\right]
\]
\[
\pm 4\Omega_{1}^2 k_u k_d \frac{1-\exp{[-(k_d+k_u)t_2]}}{k_d+k_u} {\bigg\}}
\]
where the upper (lower) sign is for $R_i$ ($R_{ii}$). Since the evolution in the excited
$|ee\rangle \rangle$ and in the ground $|gg\rangle \rangle$ state is the same
for stochastic jumps (where the bath is not affected by the system) we have 
\begin{equation}
R_{iii}(\omega_3,t_2,\omega_1)=R_{i}(\omega_3,t_2,\omega_1); \quad \quad R_{iv}(\omega_3,t_2,\omega_1)=R_{ii}(\omega_3,t_2,\omega_1)
\label{samefeynman}
\end{equation}
We further have for $R_{v}$ and $R_{vi}$:
\[
R_{\alpha}(\omega_3,t_2,\omega_1)=-
\frac{1}{(\omega_1\mp\Omega_{0})^2-\Omega_{1}^2+i(\omega_1\mp\Omega_{0})(k_d+k_u)\pm i\Omega_{1}(k_d-k_u)}
\]
\[
\times\frac{1}{(\omega_3-\Omega_{0}-\Delta_0)^2-(\Omega_{1}+\Delta_1)^2+i(\omega_3-\Omega_{1}- \Delta_1)(k_d+k_u)+ i(\Omega_{1}+\Delta_1)(k_d-k_u)}
\]
\[
\times \frac{1}{k_u+k_d} \bigg\{ (k_u+k_d)^3-\left[(\omega_3-\Omega_{0}-\Delta_0)(\omega_1\mp\Omega_{0})\pm(\Omega_{1}+\Delta_1)\Omega_{1}\right](k_u+k_d)
\]
\[
+\left[(\omega_1\mp \Omega_{1})(\Omega_{1}+\Delta_1)\pm (\omega_3-\Omega_{0}-\Delta_0)\Omega_{1}\right](k_d-k_u)
\]
\[
-i\left[(\omega_3-\Omega_{0}-\Delta_0+\omega_1\mp \Omega_{0})(k_u+k_d)^2
-(\Omega_{1} +\Delta_1\pm\Omega_{1})(k_d^2-k_u^2)\right]
\]
\[
\pm 4(\Omega_{1}(\Omega_{1}+\Delta_1) k_u k_d \frac{1-\exp{[-(k_d+k_u)t_2]}}{k_d+k_u} {\bigg\}}
\]
where the upper (lower) sign is for $R_v$ ($R_{vi}$).

\subsection{High temperature spin dynamics}
In the high temperature limit we set $k_u=k_d\equiv k$.
We first analyze the $R_{i}, R_{ii}$ contributions. Here
\begin{equation}
{\mathcal G}_{eg,eg}(t)=
e^{-(k+ i\Omega_0) t} \left(\begin{array}{cc}\cosh{(\eta t)}-\frac{i\Omega_1}{\eta}
\sinh{(\eta t)} &\frac{k}{\eta}\sinh{(\eta t)} \\ \frac{k}{\eta}\sinh{(\eta t)}
&\cosh{(\eta t)}+\frac{i\Omega_1}{\eta}\sinh{(\eta t)}\end{array} \right)
\end{equation}
and
\[
{\mathcal G}_{ge,ge}(t)={\mathcal G}_{eg,eg}^*(t)
\]
where $\eta=\sqrt{k^2-\Omega_1^2}$. When $\Omega_1>k$ we use analytic continuation
(e.g. $cosh{(ix)}=\cos{x}$ and $\sinh{(ix)}=i\sin{x}$) and
after some algebra we get for $\alpha=i,ii$ \cite{zhao}
\[ 
R_{\alpha}(t_3,t_2,t_1)= \frac{e^{-k(t_1+t_3)} e^{-i\Omega(t_3 \pm t_1)}}{2}
\]
\[
\times \left[
\left(1+\frac{k^2}{\eta^2}\right)\cosh{\left[\eta(t_1+t_3)\right]}
+\left(1-\frac{k^2}{\eta^2}\right)\cosh{\left[\eta(t_1-t_3)\right]}
+\frac{2k}{\eta}\sinh{\left[\eta (t_1+t_3)\right]}\right]
\]
\begin{equation}
\mp  e^{-k(t_1+2t_2+t_3)} e^{-i\Omega(t_3 \pm t_1)} \frac{\omega_0^2}{2\eta^2}
\left[\cosh{\left[\eta(t_1+t_3)\right]}-\cosh{\left[\eta(t_1-t_3)\right]}\right]
\end{equation}
where the upper (lower) sign stands for $R_i$ ($R_{ii}$).
Eq. (\ref{mixdomain} then gives
\[
R_{\alpha}(\omega_3,t_2,\omega_1)=
\frac{1}{(\omega_1\mp\Omega_0)^2-\Omega_{1}^2+2i(\omega_1\mp\Omega_{0})k}
\]
\[
\times \frac{1}{(\omega_3-\Omega_{0})^2-\Omega_{1}^2+2i(\omega_3-\Omega_{0})k}
\bigg\{ 4k^2-(\omega_3-\Omega_{0})(\omega_1\mp\Omega_{0})\mp\Omega_{1}^2
\]
\[
-2i(\omega_3-\Omega_{0}+\omega_1\mp\Omega_{0})k
\pm \Omega_{1}^2 \left[1-\exp{(-2k t_2)}\right] {\bigg\}}
\]
 
We further have $R_{iii}=R_{i}, \quad R_{iv}=R_{ii}$.
For $\alpha=v,vi$ we get
\[
R_{\alpha}(\omega_3,t_2,\omega_1)=
\frac{1}{(\omega_1\mp\Omega_{0})^2-\Omega_{1}^2+2i(\omega_1\mp\Omega_{0})k}
\]
\[
\times\frac{1}{(\omega_3-\Omega_{0}-\Delta_0)^2-(\Omega_{1}+\Delta_1)^2+2i(\omega_3-\Omega_{1}- \Delta_1)k}
\]
\[
\times  \bigg\{ 4k^2-\left[(\omega_3-\Omega_{0}-\Delta_0)(\omega_1\mp\Omega_{0})\pm(\Omega_{1}+\Delta_1)\Omega_{1}\right]
\]
\[
-2i\left[(\omega_3-\Omega_{0}-\Delta_0+\omega_1\mp \Omega_{0})k\right]
\pm \Omega_{1}(\Omega_{1}+\Delta_1) \left(1-\exp{[-2kt_2]}\right) {\bigg\}}
\]
where the  upper (lower) sign corresponds to $R_v$ ($R_{vi}$).

\subsection{Slow spin dynamics}
When assume $\Omega_1>>k_u,k_d$ 
no spin jumps occur during the  $t_1$,and $t_3$ intervals.
Then
\[ {\mathcal G}_{\nu \nu',\nu_1 \nu_1'}(\omega)=\delta_{\nu \nu_1}\delta_{\nu' \nu_1'}
\left(\begin{array}{cc} k_d +i(
\epsilon_0^{(\nu )}-\epsilon_0^{(\nu')} +\epsilon_1^{(\nu )}-\epsilon_1^{(\nu')}-\omega) & 0\\
0& k_u +i(
\epsilon_0^{(\nu )}-\epsilon_0^{(\nu')} -\epsilon_1^{(\nu )}+\epsilon_1^{(\nu')}-\omega)
\end{array} \right)^{-1}
\]
The Green's function for $t_2$ is still given by Eq. (\ref{decoupled}).

In the vicinity
of the peaks (Fig 2a) at e.g. $\omega_1\approx\Omega_0+\Omega_1$,
$\omega_3\approx\Omega_0-\Omega_1$  we have
\[
R_{i}(\omega_3,0,\omega_1)\propto \frac{1}{k_u-i(\omega_{1}-\Omega_0 -\Omega_1)}
\frac{1}{k_d-i(\omega_3-\Omega_0+ \Omega_1)}
\]
\[
=\frac{k_d k_u -(\omega_{1}-\Omega_0 -\Omega_1)(\omega_3-\Omega_0+ \Omega_1) +i\left[
k_d(\omega_{1}-\Omega_0 -\Omega_1)+k_u(\omega_3-\Omega_0+ \Omega_1)\right]
}{
(k_u^2+(\omega_1-\Omega_0-\Omega_1)^2)(k_d^2+(\omega_3-\Omega_0+\Omega_1)^2)}
\]
\[
R_{ii}(\omega_3,0,\omega_1)\propto
\frac{1}{k_u-i(\omega_{1}+\Omega_0+\Omega_1)}
\frac{1}{k_d-i(\omega_3-\Omega_0+ \Omega_1)}
\]
\[
=\frac{k_d k_u -(\omega_{1}+\Omega_0 +\Omega_1)(\omega_3-\Omega_0+ \Omega_1) +i\left[
k_d(\omega_{1}+\Omega_0 +\Omega_1)+k_u(\omega_3-\Omega_0+ \Omega_1)\right]
}{
(k_u^2+(\omega_1+\Omega_0+\Omega_1)^2)(k_d^2+(\omega_3-\Omega_0+\Omega_1)^2)}
\]
In the absorptive signal (Eq.(\ref{2DIRdef}))
the dispersive parts (the second term in nominator) cancel out and
the  signal is given by simple lorentzian linear
response peaks
\begin{equation}
S_A(\omega_3,0,\omega_1)\propto \frac{4\mu_{eg}^4 k_d k_u}{
\hbar^3(k_u^2+(\omega_1-\Omega_0-\Omega_1)^2)(k_d^2+(\omega_3-\Omega_0+\Omega_1)^2)}
=\hbar^{-1}  W(\omega_1) W(\omega_3)
\label{slow_peak_shapes}
\end{equation}

\subsection{Fast spin fluctuations; motional narrowing}
In the $k_u,k_d>>\Omega_1$ limit we get
\[
{\mathcal G}_{eg,eg}(\omega)
=\frac{-1}{(\omega-\Omega_0)^2+i(\omega-\Omega_0)(k_d+k_u)}
\left(\begin{array}{cc}k_u -i(\omega - \Omega_0) &k_u \\
k_d&k_d -i(\omega -\Omega_0)\end{array} \right)
\]
\[
{\mathcal G}_{ge,ge}(\omega)
={\mathcal G}_{eg,eg}(-\omega)^*
\]
The ine shapes is insensitive to $t_2$ delay one can set
\[ {\mathcal G}_{ee,ee}(t_2)={\mathcal G}_{gg,gg}(t_2)=
\frac{1}{k_u +k_d}\left(\begin{array}{cc}k_u  &k_u \\
k_d&k_d \end{array} \right)=\frac{1}{k_u +k_d}
\left(\begin{array}{c}k_u \\ k_d\end{array} \right)
\left(\begin{array}{cc}1  &1 \end{array} \right)
\]
 The response functions may now be factorized into products 
of the linear response functions (Eq. (\ref{prefactorization})).

\section{The Cumulant expansion for the spectral diffusion model}
\label{sec:Cumulant} 
Neglecting spin dynamics by
setting $\Delta_1=\Delta_3=0,\Omega_1=\Omega_3=0$,  the response functions may be
calculated by the second order cumulant expansion
\cite{principles,mukamelcum} which
represents stochastic Gaussian fluctuations.

The response function may be expressed in terms of the following four--point
correlation functions \cite{abramavicius}:
\begin{eqnarray}
F_1( \tau_4,\tau_3,\tau_2,\tau_1)&\equiv& \langle B(\tau_4) B^{\dagger}(\tau_3)
B(\tau_2) B^{\dagger}(\tau_1)\rangle
\nonumber \\
F_2( \tau_4,\tau_3,\tau_2,\tau_1)&\equiv& \langle B(\tau_4) B(\tau_3)
B^{\dagger}(\tau_2) B^{\dagger}(\tau_1)\rangle/2 .
\end{eqnarray}

We then have \cite{abramavicius,ravi,meiereee}:
\begin{eqnarray}
R_{i}(t_3,t_2,t_1)&=& F_1(t_1,t_1+t_2,t_1+t_2+t_3,0)
\nonumber \\
R_{ii}(t_3,t_2,t_1)&=& F_1(0,t_1+t_2,t_1+t_2+t_3,t_1)
\nonumber \\
R_{iii}(t_3,t_2,t_1)&=& F_1(t_1+t_2+t_3,t_1+t_2,t_1,0)
\nonumber \\
R_{iv}(t_3,t_2,t_1)&=& F_1(0,t_1,t_1+t_2+t_3,t_1+t_2)
\nonumber \\
R_{v}(t_3,t_2,t_1)&=&  F_2(t_1, t_1+t_2+t_3,t_1+t_2,0)
\nonumber \\
R_{vi}(t_3,t_2,t_1)&=& F_2(0,t_1+t_2+t_3,t_1+t_2,t_1)
\end{eqnarray}

Here:
\begin{eqnarray}\label{eq:f_simplified}
F_1(\tau_4,\tau_3,\tau_2,\tau_1)&= &
                     \exp [ i(-\Omega_{0}\tau_4+\Omega_{0}\tau_3-
          \Omega_{0}\tau_2+\Omega_{0}\tau_1)
           -f_1(\tau_4,\tau_3,\tau_2,\tau_1) ]
\end{eqnarray}
with:
\begin{eqnarray}
f_{1}(\tau_{1},\tau_{2},\tau_{3},\tau_{4})= & & g_{ee}(t_{21})+
g_{ee}(t_{43})+ g_{ee}(t_{32})+ g_{ee}(t_{41})-
                                  \nonumber \\
&& -g_{ee}(t_{31})-g_{ee}(t_{42}).\nonumber
\end{eqnarray}
where $t_{ij}\equiv \tau_i-\tau_j $ and
\begin{eqnarray}
F_2(\tau_4,\tau_3,\tau_2,\tau_1)&=&
                    \exp [ i(-\Omega_{0}\tau_4-(\Omega_{0}+\Delta_0)\tau_3+
          (\Omega_{0}+\Delta_0)\tau_2+\Omega_{0}\tau_1)
           -f_2(\tau_4,\tau_3,\tau_2,\tau_1) ].
\end{eqnarray}
with
\begin{eqnarray}
f_{2}(\tau_{1},\tau_{2},\tau_{3},\tau_{4})= & & g_{ee}(t_{21})+
g_{ff}(t_{32})+ g_{ee}(t_{43})- g_{ef}(t_{21})-
                                  \nonumber \\
&& -g_{ef}(t_{32})+g_{ef}(t_{31}) + g_{ee}(t_{32})+
g_{ee}(t_{41})-
                                  \nonumber \\
&& -g_{ee}(t_{31})- g_{ee}(t_{42})-g_{fe}(t_{32})-g_{fe}(t_{43})
+g_{fe}(t_{42}).
\end{eqnarray}
 $g_{ab}$ are the line broadening functions \cite{principles,abramavicius}
\begin{eqnarray}
g_{ee}(t)&=&\Omega_2^2\left[ \exp{\left( -\Lambda t\right)}+\Lambda t -1\right]
\nonumber \\
g_{ef}(t)&=&(\Omega_2+\Delta_2)\Omega_2 \left[ \exp{\left( -\Lambda t\right)}+\Lambda t -1\right]
\nonumber \\
g_{ff}(t)&=&(\Omega_2+\Delta_2)^2\left[ \exp{\left(
-\Lambda t\right)}+\Lambda t -1\right] \label{linebroad}
\end{eqnarray}
Note that $R_{i}=R_{iii}$, and $R_{ii}=R_{iv}$ as expected
for  stochastic models of frequency fluctuations .

\section{Spectrum of the Fokker-Planck equation}
\label{sec:Spectrum}
The eigenvectors of the Fokker-Planck operator Eq. (\ref{fp}) with eigenvalue  $-\alpha\Lambda$
are given by\cite{risken}
\begin{equation}
\label{herm}
|\alpha\rangle_Q=\frac{\exp{\left[-\left(Q^2 / 2\right)\right]}}{2^n \sqrt{2\pi}n!}
H_{\alpha} \left(\frac{Q}{\sqrt{2}}\right); \quad \quad \alpha =0, 1, 2, \ldots 
\end{equation}
where $H_{\alpha}$ is the Hermite polynomial
\[
H_{\alpha}(x)=(-1)^n e^{x^2}\frac{d^{\alpha}}{dx^{\alpha}}e^{-x^2}
\]
 The matrix representation of bath
densities and evolution matrices refer to to the basis 
$\{|\alpha\rangle_Q\}$. 
The matrix elements of ${\mathcal L}_Q$ are
\begin{equation}
 \left({\mathcal L}_Q\right)_{\beta,\alpha}=- \alpha\Lambda \delta_{\alpha,\beta}
\end{equation}

We use the recurrence relation:
\begin{equation}
\label{linherm}
 Q H_{\alpha}(Q)= \frac{H_{\alpha+1}(Q)}{2} +nH_{\alpha-1}(Q)
\end{equation}
The Q coordinate is then represented by the tridiagonal matrix
(Eq. (\ref{lincoup})).
\begin{equation}
\label{lincoup}  
[Q]_{\beta \alpha}=\beta \sqrt{2}\delta_{\beta,\alpha+1} 
+ \frac{1}{\sqrt 2} \delta_{\beta,\alpha-1}
\end{equation}

\section{Matrix Continued-fraction solution of the Fokker-Planck equation}
\label{sec:Continued_fraction}

The complete Liouville superoperator is a
 $18\times 18$ matrix in the joint vibrational
and spin space.

The  $Q$ variable is tridiagonal in the Fokker-Planck eigenbasis
and the complete Liouvillean may be thus arranged in the tridiagonal block structure in the Brownian coordinate variable
\begin{equation}
 i\omega+\hat{\mathcal L}=
 \left(\begin{array}{cccccc}{\mathcal Q}_0 & {\mathcal Q}_0^+ & 0 &0&\ldots&\dots\\{\mathcal Q}_1^-&{\mathcal Q}_1&{\mathcal Q}_1^+&0&\dots &\ldots \\  \ldots &\ldots &\ldots &\ldots &\ldots &\ldots \\  0 &\ldots &{\mathcal Q}_n^-&{\mathcal Q}_n&{\mathcal Q}_n^+&\ldots\\  \ldots &\ldots &\ldots &\ldots &\ldots &\ldots
\end{array} \right)
\label{trid}
\end{equation}

For Eq.(\ref{totalSLE}) the blocks are:
\[
{\mathcal Q}_n=i\omega+\hat{\mathcal L}_S+\hat{\mathcal L}_{SV} -n \Lambda\hat{1}
\]
\[
{\mathcal Q}_n^- = \frac{1}{\sqrt{2}}  \hat{\mathcal L}_{QS}
 \]
\[
{\mathcal Q}_n^+ =\sqrt{2} (n+1)\hat{\mathcal L}_{QS}
\]

The SLE may  be solved in the frequency-domain using a matrix continued fraction \cite{risken,jansen1,tanimura1}
The Green's function is given by the  inverse  (Eq.\ref{Green_function_frequency}) 
of the tridiagonal matrix Eq.(\ref{trid}). Starting with
\[
 \left[ i\omega +\hat{\mathcal L} \right]{\mathcal G}(\omega)=-\hat{1}
\]
For the off- diagonal $n \neq m$ elements we have
\begin{equation}
\label{off}
{\mathcal Q}_n^-{ \mathcal G}_{n-1,m}+{\mathcal Q}_n { \mathcal G}_{n,m}+{\mathcal Q}_n^+{ \mathcal G}_{n+1,m}=0
\end{equation}
The diagonal $n=m$ elements are
\begin{equation}
\label{on}
{\mathcal Q}_n^-{ \mathcal G}_{n-1,n}+{\mathcal Q}_n { \mathcal G }_{n,n}+{\mathcal Q}_n^+{ \mathcal G}_{n+1,n}=-1
\end{equation}
The recursion relation Eq. (\ref{off}) is independent of the index $m$.
Consequently  we can introduce the matrices ${\mathcal S}^+, {\mathcal S}^-$
\begin{equation}
{ \mathcal G}_{n \pm 1,b}= {\mathcal S}_n^{\pm }{ \mathcal G}_{n,m}
\label{connectmat}
\end{equation}
Using Eq. (\ref{off}) these matrices may be solved iteratively.
\begin{equation}
\label{iter}
S_n^{\pm}=\frac{-1}{{\mathcal Q}_{n\pm 1}+ {\mathcal Q}_{n\pm 1}^{\pm} {\mathcal S}_{n\pm 1}^{\pm }}{\mathcal Q}_{n\pm 1}^{\mp}
\end{equation}
Combined with Eq. (\ref{on}) we obtain for the diagonal terms:
\begin{equation}
{\mathcal G}_{n,n}=\frac{-1}{{\mathcal Q}_n^-{ \mathcal S}_n^{-}+{\mathcal Q}_n +{\mathcal Q}_n^+{ \mathcal S}_n^+}
\label{ondiag}
\end{equation}
while the off-diagonal
${\mathcal{G}}(\omega)_{nm}$ are obtained from Eq.(\ref{iter}) and (\ref{connectmat}).
For $n>m$
\begin{equation}
{\mathcal{G}}_{n,m}(\omega)=
{\mathcal{S}}_{n-1}^{+}(\omega) {\mathcal{S}}_{n-2}^{+}(\omega)\cdots {\mathcal{S}}_{m}^{+}(\omega) {\mathcal{G}}_{m,m}(\omega)
\end{equation}
and for $n<m$
\begin{equation}
{\mathcal{G}}_{n,m}(\omega)=
{\mathcal{S}}_{n+1}^{-}(\omega) {\mathcal{S}}_{n+2}^{-}(\omega)\cdots {\mathcal{S}}_{m}^{-}(\omega)  {\mathcal{G}}_{m,m}(\omega)
\end{equation}

The full Green's function ${\mathcal{G}}(\omega)$
can be calculated by using Eq. (\ref{iter}) to find
the connection matrices ${\mathcal{S}}^{\pm}$.
We note that ${\mathcal{S}}^{-}_0$ is zero and truncate the
recurrence relation for ${\mathcal{S}}^{+}$ at some level $n$
by setting ${\mathcal{S}}^{+}_n$ equal to zero.
The matrix elements ${\mathcal{G}}(\omega)_{mm}$ can
then be obtained using Eq. (\ref{ondiag}).
All other matrix elements ${\mathcal{G}}(\omega)_{nm}$
are calculated from Eq. (\ref{connectmat}).

\newpage

{\bf \Large Figure captions}\newline
\begin{description}
\item{Fig 1}
Top: Pulse configuration for a three pulse cohrent experiment \newline
B0ottom: Feynman diagrams for the third order coherent response of an anharmonic vibrations.
The diagrams correspond respectively to the six terms in Eq. (\ref{resdef}). 
\item{Fig 2} (Color Online) 2D signals  $S_A$ (Eq.(\ref{2DIRdef}) for the TSJ model (Eqs. (\ref{TSJ_SLE}), (\ref{2DIRdef})), (a), static limit $\Omega_1/k_u=5$ for short delay $k_u t_2=0$, $\Delta_0=-4\Omega_1$, $k_d=k_u$, $\Delta_1=0$. \newline
(b), static limit  $\Omega_1/k_u=5$,for long delay $k_u t_2=2.0$, $\Delta_0=-4\Omega_1$, $k_d=k_u$, $\Delta_1=0$ \newline
(c), fast chemical exchange (motional narrowing) $\Omega_1/k_u=0.2$, $\Delta_0=-4\Omega_1$,
$\Delta_1=0$,$k_d=k_u$, $k_u t_2=0$.
\item{Fig 3} (Color online)  
The 2D signal $S_A$ signal (Eqs.(\ref{2DIRdef}))
including TSJ and the fast spectral diffusion: $\Omega_2/\Lambda=0.2$; $\Omega_2/\Omega_1=3$,
$k_d=k_u=0.002\Omega_1$; $\Delta_0=-4\Omega_1$, $\Omega_3=\Delta_3=\Delta_1=\Delta_2=0$; \newline (a)
$k_d t_2=0$; (b) $k_dt_2=1$.
\item{Fig 4} (Color online)
 The 2D $S_A$ signal (Eqs.(\ref{2DIRdef})).
 with TSJ and slow spectral diffusion.$\Omega_2/\Lambda=5$, $\Omega_2/\Omega_1=0.5$,
$k_d=k_u=0.002\Omega_1$, $\Delta_0=-4\Omega_1$, $\Omega_3=\Delta_3=\Delta_2=\Delta_1=0$; \newline at times a,
$\Lambda t_2=0$; b, $\Lambda t_2=5$ (i.e. $k_d t_2 =0.1$) c, $k_d t_2=1$.
\item{Fig 5} (Color online) 2D signals for  $S_I$ generated along ${\bf k_I}$ (Eq. (\ref{k_1})),  and $S_{II}$ generated along ${\bf k_{II}}$ (Eq. \ref{k_2}) 
\newline (a) $k_{II}$, $\Lambda t_2=0$
\newline (b) $k_I$, $\Lambda t_2=0$
\newline (c) $k_{II}$, $\Lambda t_2=5$
\newline (d) $k_I$, $\Lambda t_2=5$
\newline
Other parameters same as in Fig 4.  
\item{Fig 6} (Color online) 2D signal $S_A$  (Eqs.(\ref{2DIRdef})) with  spectral diffusion and TSJ.
Spectral diffusion linewidth is varied ($\Delta_2\neq 0$). Slow spectral diffusion in $eg$ state, fast diffusion in overtone $ef$: $\Omega_2/\Lambda=2$, $\Omega_2/\Omega_1=0.6$,
$k_d=k_u=0.002\Omega_1$, $\Delta_0=-4\Omega_1$, $\Delta_2=-0.75\Omega_2$, $\Omega_3=\Delta_3=\Delta_1=0$; 
\newline  (a) $\Lambda t_2=0$; (b) $\Lambda t_2=15$ (i.e. $k_d t_2 =0.1$); (c) 
$k_d t_2=1$.
\item{Fig 7} (Color online) Absorption lineshape for our model Eq.(\ref{totalSLE}) with different spectral diffusion
width in the $u$ and $d$ states ($\Omega_3\neq 0$).
 $\Omega_1=0.5fs^{-1}$ $\Lambda=0.4ps^{-1}$; $\Omega_2=0.33ps^{-1}$, $\Omega_3=-0.07ps^{-1}$
$k_d=0.125 ps^{-1}$, $k_u=0.1 ps$, $\Delta_0=-2.0ps^{-1}$,$\Delta_2=0$, $\Delta_3=\Delta_1=0$; \newline
\item{Fig 8 (Color online)}
The 2D signal  $S_A$  (Eq. \ref{2DIRdef}) for the same parameters  of Fig 7 at various time delays (a)
$t_2=0$; (b) $t_2=2 ps$;  (c) 
$t_2=10 ps$. These spectra close resemble the experimental result of \cite{fayer}.
\end{description}

\newpage
\begin{center}
\scalebox{0.60}[0.60]{\includegraphics{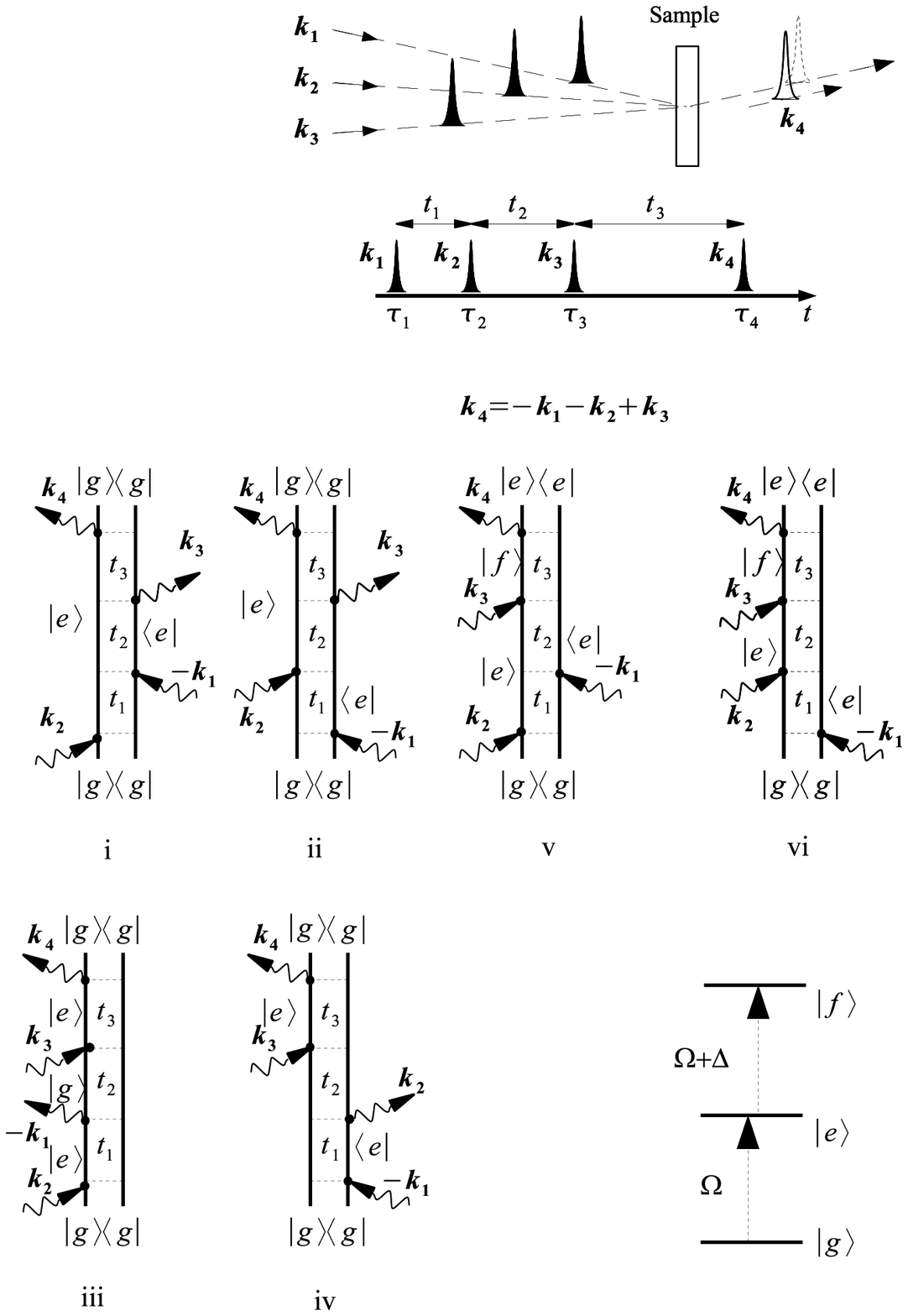}}
\end{center}
$\quad \quad \quad  \quad \quad\quad \quad \quad \quad \quad$   {\large \bf Fig 1}

\newpage
\begin{center}
\scalebox{1.20}[1.20]{\includegraphics{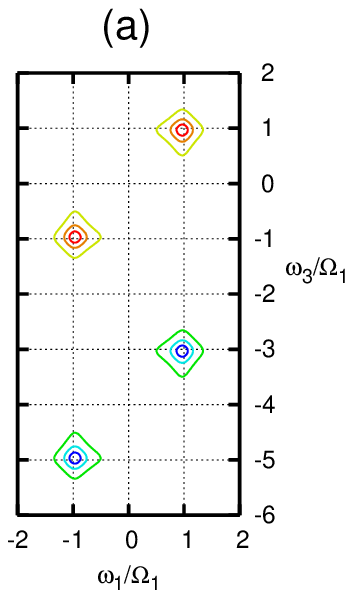}}
\scalebox{1.20}[1.20]{\includegraphics{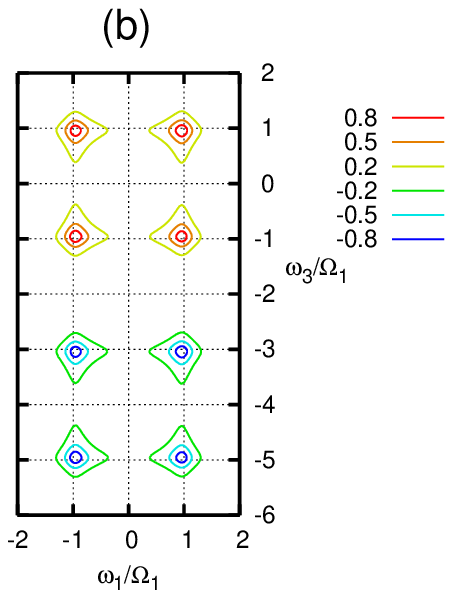}}
\scalebox{1.20}[1.20]{\includegraphics{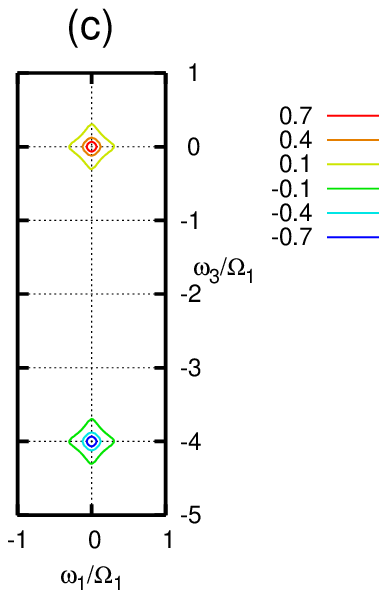}}
\vspace{85mm}
\end{center}
$\quad \quad \quad  \quad \quad\quad \quad \quad \quad \quad$   {\large \bf Fig 2}
\newpage
\begin{center}
\scalebox{1.20}[1.20]{\includegraphics{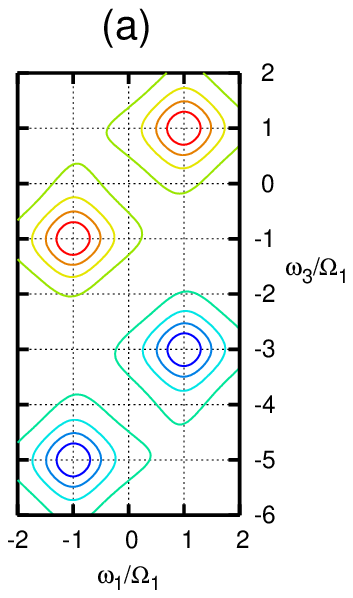}}
\scalebox{1.20}[1.20]{\includegraphics{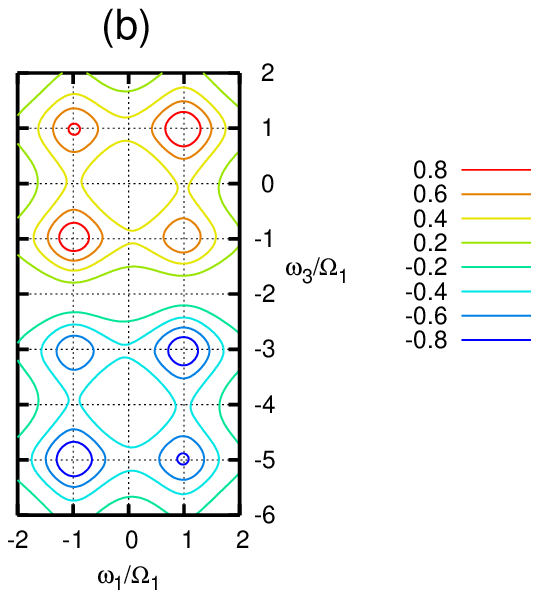}}
\vspace{95mm}
\end{center}
$\quad \quad \quad  \quad \quad\quad \quad \quad \quad \quad$   {\large \bf Fig 3}

\newpage
\begin{center}
\scalebox{1.20}[1.20]{\includegraphics{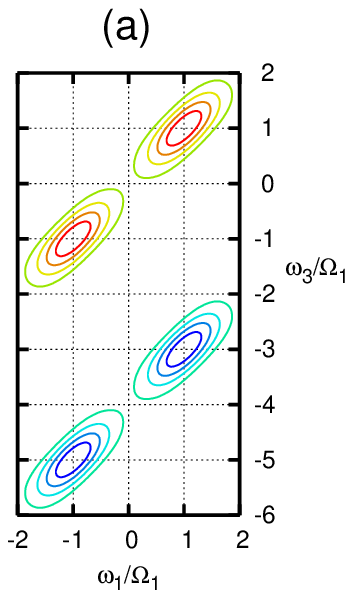}}
\scalebox{1.20}[1.20]{\includegraphics{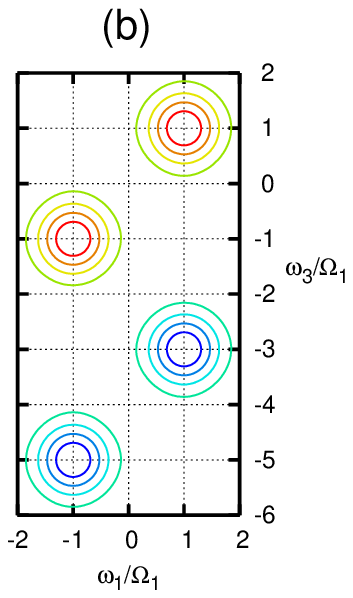}}
\scalebox{1.20}[1.20]{\includegraphics{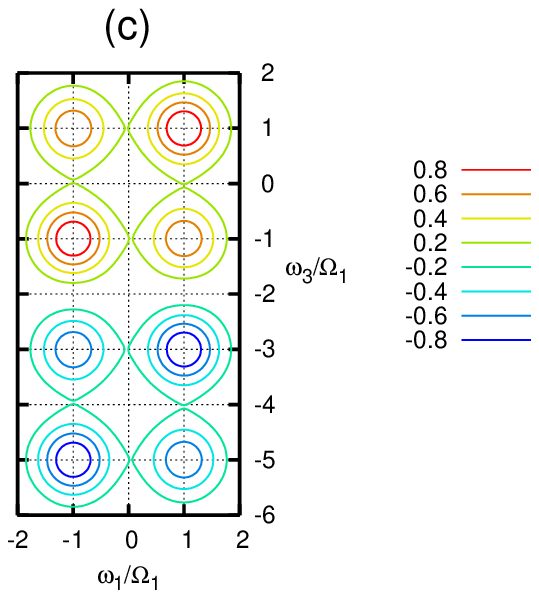}}
\vspace{95mm}
\end{center}
$\quad \quad \quad  \quad \quad\quad \quad \quad \quad \quad$   {\large \bf Fig 4}
\newpage
\begin{center}
\scalebox{1.20}[1.20]{\includegraphics{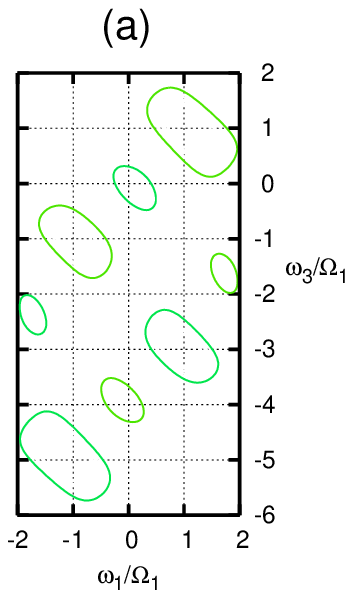}}
\scalebox{1.20}[1.20]{\includegraphics{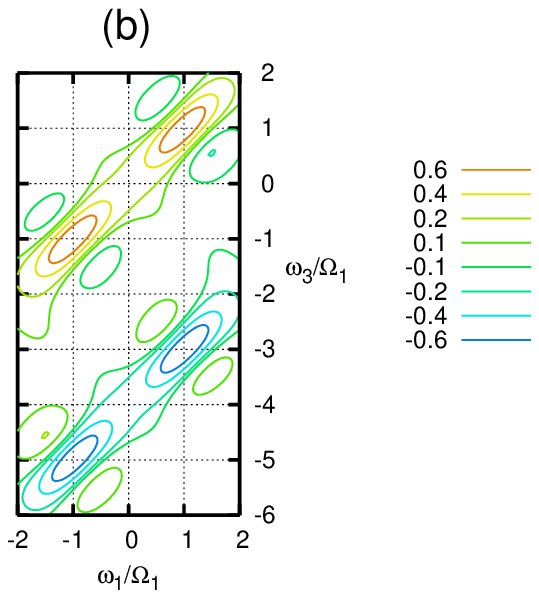}}
\scalebox{1.20}[1.20]{\includegraphics{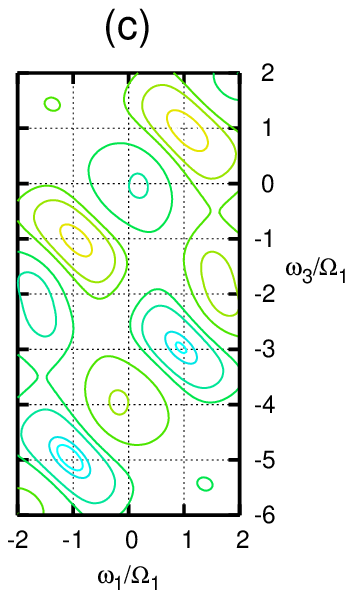}}
\scalebox{1.20}[1.20]{\includegraphics{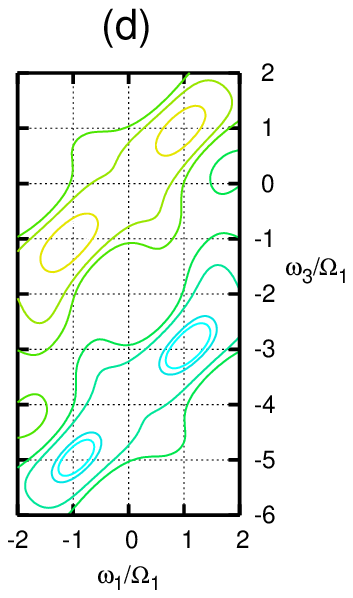}}
\vspace{55mm}
\end{center}
$\quad \quad \quad  \quad \quad\quad \quad \quad \quad \quad$   {\large \bf Fig 5}
\newpage
\begin{center}
\scalebox{1.20}[1.20]{\includegraphics{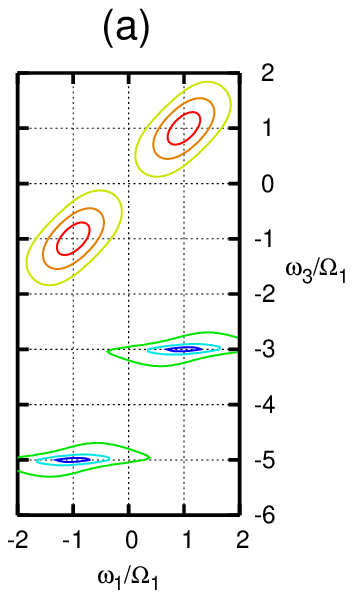}}
\scalebox{1.20}[1.20]{\includegraphics{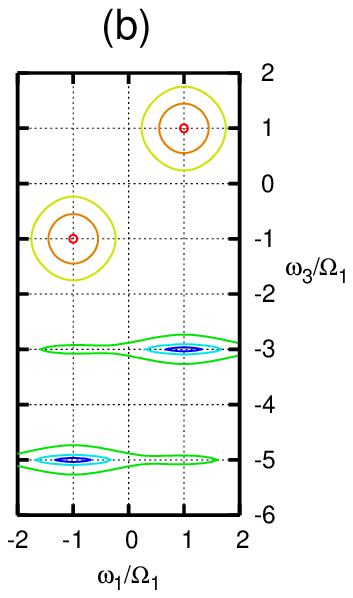}}
\scalebox{1.20}[1.20]{\includegraphics{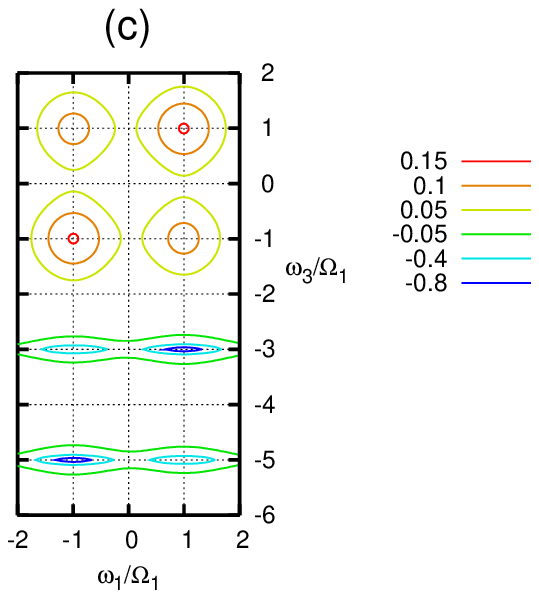}}
\vspace{95mm}
\end{center}
$\quad \quad \quad  \quad \quad\quad \quad \quad \quad \quad$   {\large \bf Fig 6}
\newpage
\begin{center}
\scalebox{1.20}[1.20]{\includegraphics{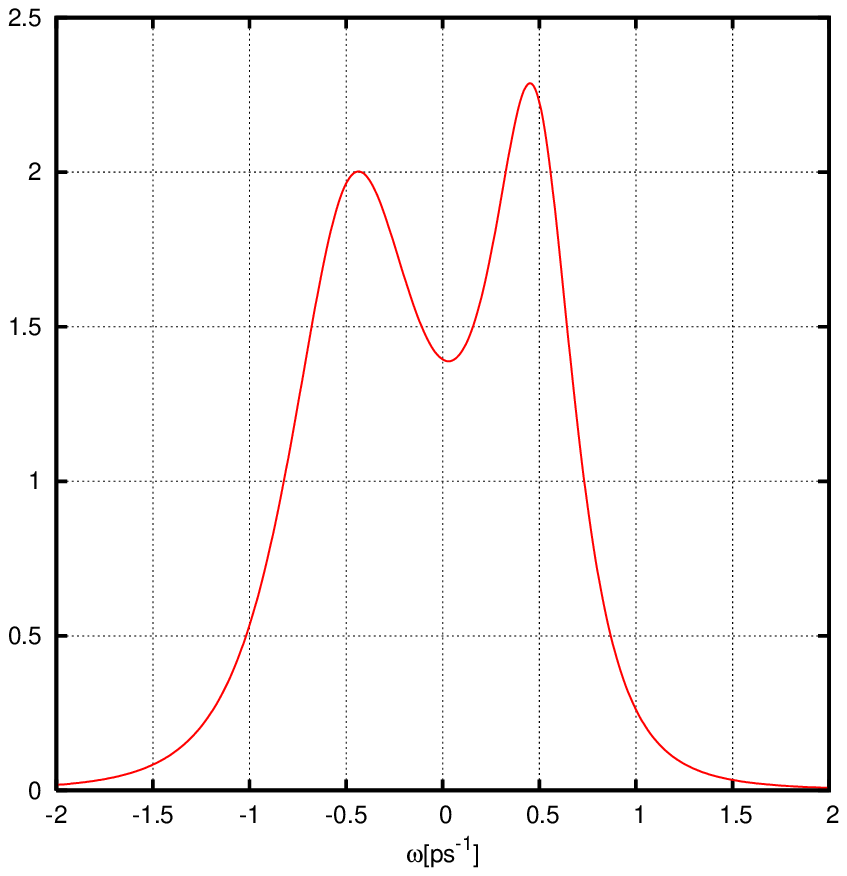}}
\vspace{15mm}
\end{center}
$\quad \quad \quad  \quad \quad\quad \quad \quad \quad \quad$   {\large \bf Fig 7}
\newpage
\newpage
\begin{center}
\scalebox{1.2}[1.2]{\includegraphics{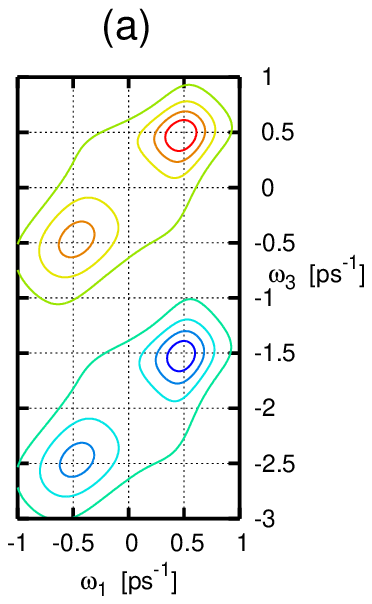}}
\scalebox{1.2}[1.2]{\includegraphics{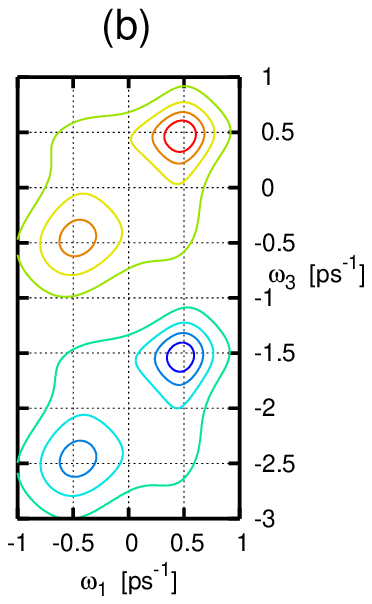}}
\scalebox{1.2}[1.2]{\includegraphics{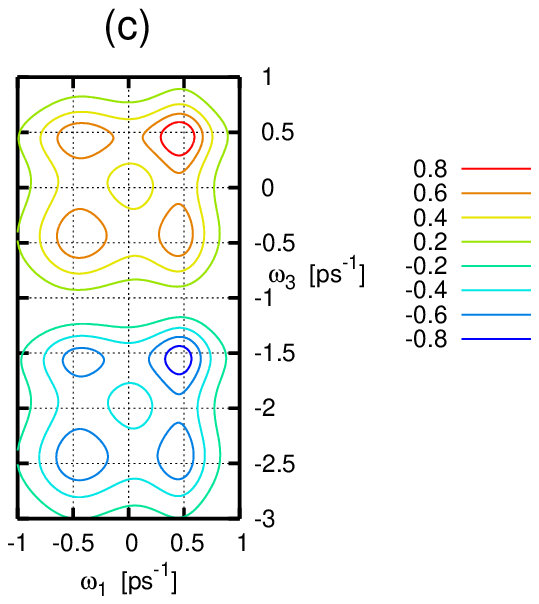}}
\vspace{95mm}
\end{center}
$\quad \quad \quad  \quad \quad\quad \quad \quad \quad \quad$   {\large \bf Fig 8}
\end{document}